\documentclass[aps,prd,12pt,notitlepage,tightenlines]{revtex4-1}
\usepackage{amsmath}
\usepackage{bm}
\usepackage{times}
\usepackage{braket}
\usepackage{color}
\usepackage{epsfig}
\usepackage{slashed}
\usepackage{hyperref}
\newcommand{\beq}{\begin{eqnarray}}
\newcommand{\eeq}{\end{eqnarray}}
\newcommand{\non}{\nonumber\\ }

\newcommand{\acp}{ {\cal A}_{CP} }
\newcommand{\cala}{ {\cal A} }
\newcommand{\cals}{ {\cal S} }
\newcommand{\calh}{ {\cal H} }

\newcommand{\mw}{ M_W}

\newcommand{\mbs}{m_{B_s} }

\newcommand{\psl}{ P \hspace{-2.4truemm}/ }
\newcommand{\nsl}{ n \hspace{-2.2truemm}/ }
\newcommand{\vsl}{ v \hspace{-2.2truemm}/ }

\newcommand{\ov}{ \overline  }

\def \cpc{ Chin. Phys. C  }

\def \epjc{ Eur. Phys. J. C }
\def \jpg{  J. Phys. G }
\def \npb{  Nucl. Phys. B }
\def \plb{  Phys. Lett. B }
\def \prd{  Phys. Rev. D }
\def \prl{  Phys. Rev. Lett.  }
\def \zpc{  Z. Phys. C }
\def \jhep{ J. High Energy Phys.  }

\definecolor{Red}{rgb}{1.,0.,0.}

\definecolor{Blue}{rgb}{0.,0.,1.}

\definecolor{nicered}{rgb}{0.7,0.1,0.1}
\definecolor{nicegreen}{rgb}{0.1,0.5,0.1}

\hypersetup{colorlinks,citecolor=nicegreen,linkcolor=nicered}
\begin{document}
%%%%%%%%%%%%%%%%%%%%%%%%%%%%%%%%%%%%%%%%%%%%%%%%%%%%
%%
\title{$\bar{B}^0_s \to K\pi,KK$ decays and effects of the next-to-leading order contributions}
\author{Jing-Jing Wang, Dong-Ting Lin, Wen Sun, Zhong-Jian Ji, Shan Cheng, and Zhen-Jun Xiao
\email[Electronic address:]{xiaozhenjun@njnu.edu.cn} }
\affiliation{Department of Physics and Institute of
Theoretical Physics, Nanjing Normal University, Nanjing, Jiangsu 210023, P.R.China }
\date{\today}
\begin{abstract}
By employing the perturbative QCD(pQCD) factorization approach, we calculate 
the branching ratios and CP violating asymmetries of the four $\bar{B}_s^0 \to K\pi$ 
and $K K $ decays, with the inclusion of all known next-to-leading order (NLO) 
contributions. We find numerically that
(a) the NLO contribution can interfere with the LO part constructively or destructively 
for different decay modes;
(b) the NLO contribution leads to a $22\%$ decrease for 
$Br(\bar{B}^0_s \to K^+ \pi^-)$, but $\sim 50\%$ enhancements to other three
considered $\bar{B}_s$ decays, and therefore play an important role in 
interpreting the measured values of the branching ratios; and 
(c) for both $\bar{B}_s^0 \to K^+ \pi^-$ and $\bar{B}_s^0 \to K^+K^- $ decays, 
the NLO pQCD predictions for the direct and mixing induced CP-violating asymmetries 
agree very well with the measured values in both the sign and the magnitude.
\end{abstract}

\vspace{1cm}


\maketitle

%%--------------------------------------------------

\section{Introduction}

The $B$ and $B_s$ decays are very interesting phenomenologically for the precision
test of the standard model (SM) and for the searches for the signal of the
new physics beyond the SM. But the $B_s$ decays are considerably less studied
than the well-known $B_{u,d}$ decays due to the
rapid oscillations of $B_s$ mesons and the shortage of $B_s$ events collected.
Since the start of the LHC running, a lot of $B_s^0 $ events have been
collected by LHCb collaboration, and some $B_s^0 \to PP$ decays are already
observed \cite{lhcb1,lhcb2}: such as the first observation of the
direct CP violation in $B_s$ decays \cite{lhcb1}:
\beq
\acp(B_s^0 \to K^-\pi^+)=(0.27\pm 0.04(stat) \pm 0.01(syst)),\label{eq:acp-kpi}
\eeq
and the first measurement of the time-dependent CP violation in
$B_s^0 \to K^+ K^-$ \cite{lhcb2}:
\beq
C_{KK}&=&0.14\pm 0.11(stat)\pm 0.03(syst),\non
S_{KK}&=&0.30\pm 0.12(stat)\pm 0.04(syst).\label{eq:acp-skk}
\eeq

During the past decade, in fact, many charmless two-body hadronic $B_s^0\to M_2 M_3$
decays have been studied by employing the pQCD factorization approach at the LO
level \cite{bspipi,pieta,ali07} or the partial NLO level \cite{xiao08a}.
In this paper we calculate the branching ratios and CP violating asymmetries
of the $B_s^0\to K\pi$ and $K K$ decays by employing the pQCD factorization approach,
with the inclusion of all known NLO contributions.
These decay modes have also been studied, for example, by using the pQCD approach at
the LO level in Ref.~\cite{ali07}, the generalized factorization
in Ref.~\cite{chenbs99} or
by using the QCD factorization approach in Ref.~\cite{npb675,sun2003,chengbs09}.

In the pQCD factorization approach, almost all NLO contributions to $B_{u,d}\to M_2 M_3$
decays have been calculated up to now.
And it is straightforward to extend these calculations to the cases for the similar
$B_s \to M_2 M_3$ decays.
The NLO pQCD predictions for those considered decay modes proved that
the NLO contributions can play an important
role in understanding the very large $Br(B \to K \eta^\prime)$ \cite{xiao08b,fan2013}
or the so-called ``$K\pi$-puzzle" \cite{xiao2014}.
We here focus on the studies for the possible effects of the NLO contributions from
various sources: such as the QCD vertex corrections (VC), the quark-loops (QL)
and the chromo-magnetic penguins (CMP) \cite{npb675,nlo05}.
The newly known NLO twist-2 contribution \cite{prd85-074004}
and NLO twist-3 contribution to the relevant form factors \cite{cheng14a} will also
be taken into account here.
By this way, one can improve the reliability of the pQCD factorization approach
effectively.

This paper is organized as follows. In Sec.~\ref{sec:f-work}, we
give a brief review about the pQCD factorization approach and present
the LO decay amplitudes for the studied decay modes.
In Sec.~\ref{sec:nlo}, the NLO contributions from different sources
are evaluated analytically. We calculate and show the pQCD predictions for the
branching ratios and CP violating asymmetries of $B_s^0 \to K \pi$ and $KK$ decays
in Sec.~IV. The summary and some discussions are included in the
final section.

\section{ Theoretical framework and LO decay amplitudes }\label{sec:f-work}

\subsection{ Outlines of the pQCD approach}\label{sec:2-1}

We consider the $B$ meson at rest for simplicity.
Using the light-cone coordinates, we define the $B_s^0$ meson with momentum $P_1$,
the emitted meson $M_2$ with momentum
$P_2$ moving along the direction of $n=(1,0,{\bf 0}_{\rm T})$ and the recoiled
meson $M_3$ with momentum $P_3$ in the direction of
$v=(0,1,{\bf 0}_{\rm T})$. Here we also use $x_i$ to denote the momentum fraction
of anti-quark in each meson:
\beq
P_1 &=& \frac{\mbs}{\sqrt{2}} (1,1,{\bf 0}_{\rm T}), \quad
P_2 =\frac{\mbs}{\sqrt{2}}(1-r_3^2,r^2_2,{\bf 0}_{\rm T}), \quad
P_3 =\frac{\mbs}{\sqrt{2}} (r_3^2,1-r^2_2,{\bf 0}_{\rm T}), \\
k_1 &=& \frac{\mbs}{\sqrt{2}}\left (x_1,0,{\bf k}_{\rm 1T}\right), \quad
k_2 =\frac{\mbs}{\sqrt{2}}\left (x_2(1-r_3^2),x_2 r^2_2,{\bf k}_{\rm 2T} \right ),\non
k_3&=& \frac{\mbs}{\sqrt{2}}\left (x_3 r_3^2,x_3(1-r_2^2),{\bf k}_{\rm 3T} \right),
\eeq
where $r_i=m_i/\mbs$ with $m_i=m_\pi$ or $m_K$ here.
When the light pion and kaon are the final state mesons,
$r_i^2 < 0.01$ and can be neglected safely.
The integration over the small
components $k_1^-$, $k_2^-$, and $k_3^+$ will lead conceptually to the decay amplitudes,
\beq
{\cal A}(B_s \to M_2M_3 ) &\sim &\int\!\! d x_1 d x_2 d x_3 b_1 d b_1 b_2 d b_2 b_3 d b_3 \non &&
\cdot \mathrm{Tr} \left [ C(t) \Phi_{B_s}(x_1,b_1) \Phi_{M_2}(x_2,b_2)
\Phi_{M_3}(x_3, b_3) H(x_i, b_i, t) S_t(x_i)\, e^{-S(t)} \right ],
\quad \label{eq:a2}
\eeq
where $b_i$ is the conjugate space coordinate of $k_{iT}$.
In the above equation, $C(t)$ is the Wilson coefficient evaluated at scale $t$.
The functions $\Phi_{B_s}$, $\Phi_{M_2}$ and $\Phi_{M_3}$ are the wave functions
of the initial $B_s$ meson and the final-state meson $M_2$ and $M_3$ respectively.
The hard kernel  $H(k_1,k_2,k_3,t)$ describes the four-quark operator and the
spectator quark connected by  a hard gluon whose $q^2$ is in the order
of $\bar{\Lambda} \mbs$. The jet function $S_t(x_i)$ in Eq.(\ref{eq:a2})
is one of the two kinds of Sudakov form factors relevant for the $B_s$ decays
considered, which come from the threshold resummation over the large double
logarithms ($\ln^2 x_i$) in the end-point region.
The function $ e^{-S(t)}$ is the second kind of the Sudakov form factors.
The Sudakov form factors suppress effectively the soft dynamics
at the end-point region \cite{li2003}.

For the studied $\bar{B}_s^0 \to K\pi, KK$ decays, the corresponding weak
effective Hamiltonian can be written as \cite{buras96}
\beq
{\cal H}_{eff} &=& \frac{G_{F}}{\sqrt{2}}     \Bigg\{ V_{ub} V_{uq}^{\ast} \Big[
 C_{1}({\mu}) O^{u}_{1}({\mu})  +  C_{2}({\mu}) O^{u}_{2}({\mu})\Big]
  -V_{tb} V_{tq}^{\ast} \Big[{\sum\limits_{i=3}^{10}} C_{i}({\mu}) O_{i}({\mu})
  \Big ] \Bigg\} + \mbox{h.c.} ,
\label{eq:heff}
\eeq
where $q=d,s$, $G_{F}=1.166 39\times 10^{-5} GeV^{-2}$ is the Fermi constant,
and $V_{ij}$ is the Cabbibo-Kobayashi-Maskawa (CKM) matrix element,
$C_i(\mu)$ are the Wilson coefficients evaluated
at the renormalization scale $\mu$ and $O_i(\mu)$ are the four-fermion operators.

As usual, we treat the $B$ meson as a very good heavy-light system, and
adopt the distribution amplitude $\phi_{B_s}$ as in Ref.~\cite{ali07}
\beq
\phi_{B_s}(x,b)&=& N_{B_s} x^2(1-x)^2 \mathrm{exp} \left
 [ -\frac{M_{B_s}^2\ x^2}{2 \omega_{b}^2} -\frac{1}{2} (\omega_{b} b)^2\right],
 \label{phib}
\eeq
where the shape parameter $\omega_b$ is a free parameter and we take
$\omega_b=0.5\pm 0.05$ GeV for $B_s$ meson based on studies of lattice
QCD and light-cone sum rule\cite{li2003}, and
finally the normalization factor $N_{B_s}$ depends on the values
of $\omega_b$ and the decay constant $f_{B_s}$ and defined through the normalization
relation $\int_0^1 dx\; \phi_{B_s}(x,0)=f_{B_s}/(2\sqrt{6})$.

For the light pseudo-scalar mesons $\pi$ and $K$, their wave functions are
the same in form and can be defined as
\cite{pball98}
\beq
\Phi(P,x,\zeta)\equiv \frac{1}{\sqrt{2N_C}}\gamma_5 \left [ \psl
\phi^{A}(x)+m_0 \phi^{P}(x)+ \zeta m_0 (\nsl \vsl -1)\phi_{P}^{T}(x)\right ],
\label{eq:phi-x1}
\eeq
where $P$ and $x$ are the momentum of the light meson and the momentum fraction of the quark
(or anti-quark) inside the meson, respectively.
When the momentum fraction of the quark (anti-quark) is set to be $x$, the parameter
$\zeta$ should be chosen as $+1$ ($-1$).
The distribution amplitudes (DA's) of the light meson $M=(\pi, K)$
are adopted from Ref.~\cite{pball98,pball06}:
\beq
\phi_M^A(x) &=&  \frac{3 f_M}{\sqrt{6} } x (1-x)
    \left[1+a_1^{M}C^{3/2}_1(t)+a^{M}_2C^{3/2}_2(t)\right],\label{eq:piw1}\\
\phi_M^P(x) &=&   \frac{f_M}{2\sqrt{6} }
   \left [ 1+\left (30\eta_3-\frac{5}{2}\rho^2_{M} \right ) C^{1/2}_2(t)
   \right ], \ \
\label{eq:piw2}   \\
\phi_M^T(x) &=&  \frac{f_M(1-2x)}{2\sqrt{6} }
   \left[ 1+6\left (5\eta_3-\frac{1}{2}\eta_3\omega_3-\frac{7}{20}\rho^2_M
   -\frac{3}{5}\rho^2_M a_2^{M} \right )
   \left (1-10x+10x^2\right )\right],\quad
   \label{eq:piw3}
\eeq
with the mass ratio $\rho_M=(m_\pi/m_0^\pi,m_K/m_0^K)$ for $M=(\pi, K)$
respectively \cite{nlo05,xiao08b}. The Gegenbauer moments
$a_i^M$ and other input parameters are the same as in Ref.~\cite{ali07}:
\beq
a^\pi_1&=&0,\quad a^\pi_2=0.44^{+0.10}_{-0.20}, \quad a^K_1  = 0.17\pm 0.05,
\quad a^K_2=0.20\pm 0.06,\non
\eta_3&=&0.015, \quad \omega_3=-3.0.
\eeq
The Gegenbauer polynomials $C^{\nu}_n(t)$ in Eqs.~(\ref{eq:piw1}-\ref{eq:piw3})
can be found easily in Refs.~\cite{ali07,fan2013}. For more details about
recent progress on the wave functions of heavy and light mesons,
one can see Ref.~\cite{wu2014} and references therein.

%%%%%%%%%%%%%%%%%%%%%%%%%%%%%%%%%%%%%%%%%%%%%%%%%%%%%%%%%%%%%%%%%%%%%%%%%%%%%%%%%%%%%%%%%

\subsection{ Decay amplitudes at leading order}\label{subsec:lo}

\begin{figure}[tb]
\vspace{-5cm}
\centerline{\epsfxsize=16cm \epsffile{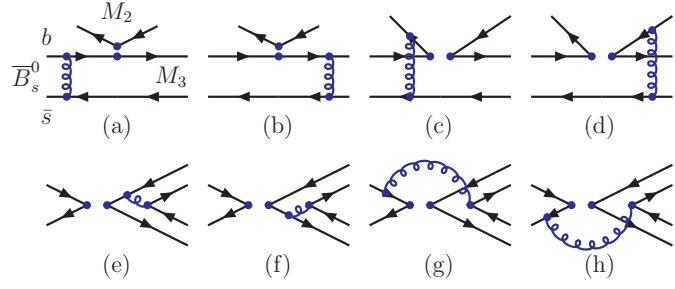}}
\vspace{-14.5cm}
\caption{Typical Feynman diagrams which may contribute at leading order
to $\bar{B}_s^0 \to K \pi, KK$ decays.} \label{fig:fig1}
\end{figure}

The four $\bar{B}_s^0 \to (K^+\pi^-,K^0\pi^0,K^+K^-,\ov{K}^0K^0)$ decays
have been studied previously in Ref.~\cite{ali07} by employing the pQCD
factorization approach at leading order.
The decay amplitudes as presented in Ref.\cite{ali07} are confirmed by our
recalculation. In this paper, we focus on the calculations of the NLO
contributions to these decays. At the leading order, the relevant Feynman diagrams
which may contribute to the $B_s^0\to K\pi, KK$ decays are  illustrated
in Fig.~\ref{fig:fig1}. For the sake of completeness, however, we firstly show the  relevant
LO decay amplitudes in this section based on our own analytical calculations.

\beq
{\cal A}(\bar{B}_s^0\to K^+ \pi^-) &=& V_{ub}V_{ud}^*\cdot
\left [ f_\pi F_{eK}\; a_1 + M_{eK}\; C_1 \right ] - V_{tb}V_{td}^*\cdot \Biggl \{ f_\pi F_{eK} \left ( a_4+a_{10}\right)
\non
&& \hspace{-1cm}
+ f_\pi F_{eK}^{P_2} \left ( a_6+a_8 \right)  + M_{eK}\left ( C_3+C_9 \right)
+ f_{B_s}F_{aK}\left ( a_4-\frac{1}{2}a_{10}\right)
\non
&& \hspace{-1cm}
+ f_{B_s} F_{aK}^{P_2}\left ( a_6-\frac{1}{2}a_8\right)+ M_{aK}\left (C_3-\frac{1}{2}C_9 \right)
+ M_{aK}^{P_1}\left ( C_5-\frac{1}{2}C_7 \right)\Biggr \},
\label{eq:das-01}
\eeq
\beq
\sqrt{2}{\cal A}(\bar{B}_s^0\to K^0\pi^0) &=&
V_{ub}V_{ud}^*\cdot  \left [ f_\pi F_{eK} a_2 + M_{eK} C_2\right]
- V_{tb}V_{td}^*\cdot \Bigl \{ - f_{B_s}F_{aK}\left ( a_4- \frac{1}{2}a_{10}\right)
\non
&& \hspace{-3cm}
-\left (  f_\pi F_{eK}^{P_2} + f_{B_s}F_{aK}^{P_2} \right ) \left (a_6-\frac{1}{2}a_8\right)
+ M_{eK}\left ( -C_3 +\frac{3}{2}C_8+\frac{1}{2}C_9+\frac{3}{2}C_{10} \right)
\non
&&  \hspace{-3cm}
+f_\pi F_{eK} \left ( -a_4-\frac{3}{2}a_7+\frac{3}{2}a_9 +\frac{1}{2}a_{10} \right)
-M_{aK}\left ( C_3- \frac{1}{2}C_9 \right)
- M_{aK}^{P_1}\left (C_5-\frac{1}{2}C_7\right) \Bigr \}, \label{eq:das-02}
\eeq
\beq
{\cal A}(\bar{B}_s^0\to K^+K^-)&=& V_{ub}V_{us}^*\cdot \left [ f_k F_{eK} a_1 + M_{eK} C_1 + M_{aK} C_2 \right]
-  V_{tb}V_{ts}^*\cdot \Biggl \{ f_k F_{eK} \left ( a_4+a_{10}\right )
\non &&  \hspace{-2cm}
+f_k F_{eK}^{P_2} \left (a_6+a_8 \right)
+ M_{eK} \left (C_3+C_9 \right ) + M_{eK}^{P_1}\left ( C_5+C_7 \right)
+f_{B_s} F_{aK}^{P_2} \left ( a_6-\frac{1}{2}a_8\right )
\non
&& \hspace{-2cm}
+ M_{aK}\left (C_3+C_4 -\frac{1}{2}C_9 -\frac{1}{2}C_{10}\right)
+ M_{aK}^{P_1} \left ( C_5-\frac{1}{2}C_7 \right)
\non
&& \hspace{-2cm}
M_{aK}^{P_2}\left (C_6-\frac{1}{2}C_8 \right)
+ \left [ M_{aK} \left ( C_4+C_{10} \right)
+ M_{aK}^{P_2} \left ( C_6+C_8 \right ) \right]_{K^+ \leftrightarrow K^-} \Biggr \},
\label{eq:das-03}
\eeq
\beq
{\cal A}( \bar{B}_s^0\to \bar{K}^0 K^0)&=&
-  V_{tb}V_{ts}^*\cdot \Biggl \{  f_k F_{eK} \left ( a_4-\frac{1}{2}a_{10} \right)
+ \left( f_k F_{eK}^{P_2} +f_{B_s}F_{aK}^{P_2} \right)
\left ( a_6-\frac{1}{2}a_8 \right) \non
&& \hspace{-3cm}
+ M_{eK}\left ( C_3-\frac{1}{2}C_{9} \right)+ \left ( M_{eK}^{P_1}+ M_{aK}^{P_1} \right)
\left( C_5- \frac{1}{2}C_{7} \right)  + M_{aK} \left ( C_3+C_4-\frac{1}{2}C_9 -\frac{1}{2}C_{10} \right)
\non
&& \hspace{-3cm}
+ M_{aK} \left ( C_4-\frac{1}{2}C_{10} \right)_{K^0 \leftrightarrow \bar{K}^0}
+ \left [ M_{aK}^{P_2}\left ( C_6-\frac{1}{2}C_8 \right) + [ K^0 \leftrightarrow \bar{K}^0]
\right] \Biggr \}, \label{eq:das-04}
\eeq
where $a_i$ is the combination of the Wilson coefficients $C_i$ the same as in Ref.~\cite{ali07}.
The nine individual decay amplitudes, such as $F_{eK}$ and $F_{eK}^{P2}$
appeared in Eqs.~(\ref{eq:das-01}-\ref{eq:das-04}), are obtained by evaluating
the corresponding Feynman diagrams in Fig.~1 analytically.
One can find the expressions for all these decay amplitudes easily in Ref.~\cite{ali07}.

\section{Next-to-leading order contributions}\label{sec:nlo}

\subsection{NLO contributions from different sources}\label{sec:vc2}

\begin{figure}[tb]
\vspace{-5cm} \centerline{\epsfxsize=16 cm \epsffile{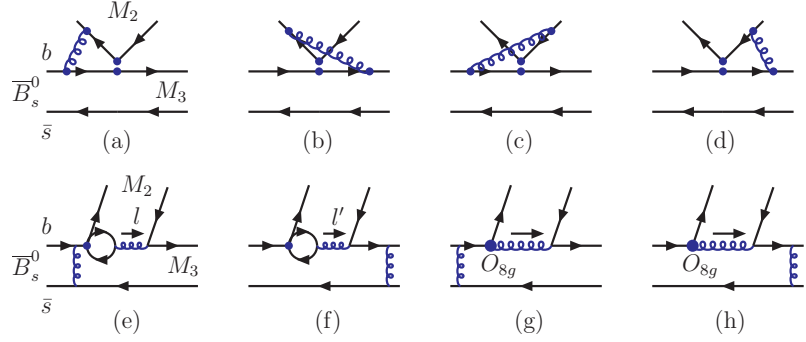}}
\vspace{-14cm}
\caption{Feynman diagrams for NLO contributions:  the vertex corrections (a-d);
the quark-loop (e-f) and the chromo-magnetic penguin contributions (g-h).}
\label{fig:fig2}
\end{figure}

\begin{figure}[tb]
\vspace{-5cm} \centerline{\epsfxsize=16 cm \epsffile{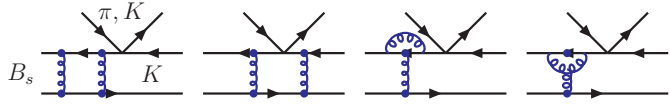}}
\vspace{-16.5cm}
\caption{The four typical Feynman diagrams, which
contribute to the form factors of $B \to M_3$ transitions
at NLO level.} \label{fig:fig3}
\end{figure}

For the considered decay modes, one should, firstly, use the NLO Wilson coefficients
$C_i(\mw)$, the NLO RG evolution matrix $U(t,m,\alpha)$ \cite{buras96} and the
$\alpha_s(t)$ at two-loop level in numerical calculations.
Secondly, one should take all the Feynman diagrams which lead to the decay amplitudes
proportional to $\alpha^2_s(t)$ in the analytical evaluations.
Such Feynman diagrams can be grouped into following classes:
\begin{enumerate}
\item
The vertex corrections, as illustrated in Figs.~\ref{fig:fig2}a-\ref{fig:fig2}d,
the same set as in the QCDF approach.

\item
The NLO contributions from quark-loops \cite{nlo05} and the chromo-magnetic penguin
operator $O_{8g}$
\cite{o8g2003},, as illustrated  in Figs.~\ref{fig:fig2}e-\ref{fig:fig2}h.

\item
The NLO contributions to the form factors of $B \to K$ transitions
\cite{prd85-074004,cheng14a}, coming from the Feynman diagrams in
Fig.~\ref{fig:fig3}.

\item
The NLO corrections to the LO hard spectator diagrams and annihilation diagrams,
as illustrated in Fig.~5 of Ref.\cite{fan2013}.

\end{enumerate}

At present, only the calculations for the NLO corrections to the
LO hard spectator and annihilation diagrams have not been completed yet.
But from the comparative studies of the LO and NLO
contributions from different sources in Ref.~\cite{fan2013,xiao2014}, we believe
that those still unknown NLO contributions in the framework of the pQCD factorization approach,
as the high order corrections to small LO contributions, are most possibly very
small in size and could be neglected safely.

The vertex corrections to the factorizable emission diagrams, as illustrated by
Figs.~2a-2d, have been calculated years ago in the QCD factorization approach
\cite{npb675,bbns99}.
For $B_s^0 \to K\pi, KK$ decays, the vertex corrections can be calculated without
considering the transverse momentum effects of the quark at the
end-point \cite{nlo05}, one can use the vertex corrections as given in
Ref.~\cite{npb675} directly.
The vertex corrections can then be absorbed into the re-definition of the
Wilson coefficients $a_i(\mu)$ by adding a vertex-function $V_i(M)$ to them.
The expressions of the vertex functions $V_{i}(M)$ can be found easily
in Refs.~\cite{npb675,nlo05}.

%%------------------------------------------

The contribution from the so-called ``quark-loops" is a kind of penguin correction
with the four quark operators insertion, as illustrated by
Fig.~\ref{fig:fig2}e and \ref{fig:fig2}f.
For the $b\to s$ transition, the effective Hamiltonian $H_{eff}^{ql}$ which describes
the contributions from the quark loops can be written as \cite{nlo05}
\beq
H_{eff}^{(QL)}&=&-\sum\limits_{q=u,c,t}\sum\limits_{q{\prime}}\frac{G_F}{\sqrt{2}}
V_{qb}V_{qs}^{*}\frac{\alpha_s(\mu)}{2\pi}C^{(q)}(\mu,l^2)\left(\bar{s}\gamma_\rho
\left(1-\gamma_5\right)T^ab\right)\left(\bar{q}^{\prime}\gamma^\rho
T^a q^{\prime}\right),
\eeq
where $l^2$ is  the invariant mass of the gluon, as illustrated by
Fig.\ref{fig:fig2}e. The expressions of the functions $C^{(q)}(\mu,l^2)$
for the loop of the $q(q=u,d,s,c,t)$ quark can be found for example
in Ref.~\cite{nlo05}.

%%---------------------------------------

The magnetic penguin is another kind penguin correction induced by the insertion of
the operator $O_{8g}$, as illustrated by Fig.2g and 2h. The corresponding weak
effective Hamiltonian contains the $b\to s g$ transition can be written as
\beq
H_{eff}^{MP} &=&-\frac{G_F}{\sqrt{2}} V_{tb}V_{ts}^*\; C_{8g}^{eff} O_{8g},
\eeq
where $O_{8g}$ is the chromo-magnetic penguin operator \cite{buras96,o8g2003} and
$C_{8g}^{eff}$ is the corresponding effective Wilson coefficient:
$C_{8g}^{eff}= C_{8g}+ C_5$ \cite{nlo05}.

%%-----------------------------------------------------

In Refs.~\cite{prd85-074004,cheng14a}, the authors calculated
the NLO twist-2 and twist-3 contributions to the form factors $f^{+,0}(q^2)$ of
the $B\to \pi $ transition. The NLO pQCD prediction for the form factor
$f^{+}(q^2)$, for example, is of the form \cite{cheng14a}
\beq
f^+(q^2)|_{\rm NLO} &=& 8 \pi m^2_B C_F \int{dx_1 dx_2} \int{b_1 db_1 b_2 db_2} \phi_B(x_1,b_1)\non
&& \times \Biggl \{ r_\pi
\left [\phi_{\pi}^{P}(x_2) - \phi_{\pi}^{T}(x_2) \right ]
\cdot \alpha_s(t_1)\cdot e^{-S_{B\pi}(t_1)}\cdot S_t(x_2)\cdot
h(x_1,x_2,b_1,b_2) \non
&&  + \Bigl [ (1 + x_2 \eta)
\left (1 + F^{(1)}_{\rm T2}(x_i,\mu,\mu_f,q^2)\; \right )
\phi_{\pi}^A(x_2)+ 2 r_\pi \left ( \frac{1}{\eta} - x_2 \right )
\phi_{\pi}^T(x_2) \non
&&  - 2x_2 r_\pi \phi_{\pi}^P(x_2) \Bigr ]
\cdot \alpha_s(t_1)\cdot e^{-S_{B\pi}(t_1)} \cdot S_t(x_2)\cdot h(x_1,x_2,b_1,b_2)\non
&&  + 2 r_{\pi} \phi_{\pi}^P(x_2)
\left (1 + F^{(1)}_{\rm T3}(x_i,\mu,\mu_f,q^2)\; \right )\non
&& \cdot \alpha_s(t_2)\cdot e^{-S_{B\pi}(t_2)}
\cdot S_t(x_2)\cdot h(x_2,x_1,b_2,b_1) \Biggr \}, \label{eq:ffnlop}
\eeq
where $\eta=1-q^2/m_B^2$ with $q^2=(P_B-P_\pi)^2$, $\mu$ ($\mu_f$) is  the
renormalization (factorization ) scale, the hard scale $t_{1,2}$ are chosen as the
largest scale of the propagators in the hard $b$-quark decay
diagrams \cite{prd85-074004,cheng14a}, the function $S_t(x_2)$ is the threshold
resummation factor adopted from Ref.~\cite{kls01}, the expressions of the hard function
$h(x_i,b_j)$ can be found in Ref.~\cite{prd85-074004,cheng14a},
and finally the factor $F^{(1)}_{\rm T2}(x_i,\mu,\mu_f,q^2)$ and
$F^{(1)}_{\rm T3}(x_i,\mu,\mu_f,q^2)$ describe the NLO twist-2 and twist-3
contribution to $f^{+,0}(q^2)$ of the $B \to \pi$ transition
respectively~\cite{prd85-074004,cheng14a}:
\beq
F^{(1)}_{\rm T2}(x_i,\mu,\mu_f,q^2)&=& \frac{\alpha_s(\mu_f) C_F}{4 \pi}
\Biggl [\frac{21}{4} \ln{\frac{\mu^2}{m^2_B}}
-(\frac{13}{2} + \ln{r_1}) \ln{\frac{\mu^2_f}{m^2_B}}
+\frac{7}{16} \ln^2{(x_1 x_2)}+ \frac{1}{8} \ln^2{x_1} \non
&& \hspace{-2cm}+ \frac{1}{4} \ln{x_1} \ln{x_2}
+ \left (- \frac{1}{4}+ 2 \ln{r_1} + \frac{7}{8} \ln{\eta} \right ) \ln{x_1}
+ \left (- \frac{3}{2} + \frac{7}{8} \ln{\eta} \right) \ln{x_2} \non
&& \hspace{-2cm}+ \frac{15}{4} \ln{\eta} - \frac{7}{16} \ln^{2}{\eta}
+ \frac{3}{2} \ln^2{r_1} - \ln{r_1}
+ \frac{101 \pi^2}{48} + \frac{219}{16} \Biggr ], \label{eq:ffnlot2}
\eeq
\beq
F^{(1)}_{\rm T3}(x_i,\mu,\mu_f,q^2)&=&\frac{\alpha_s(\mu_f) C_F}{4 \pi}
\Biggl [\frac{21}{4} \ln{\frac{\mu^2}{m^2_B}}
- \frac{1}{2}(6 + \ln{r_1}) \ln{\frac{\mu^2_f}{m^2_B}}
+ \frac{7}{16} \ln^2{x_1} - \frac{3}{8} \ln^2{x_2} \non
&&\hspace{-2cm}+ \frac{9}{8} \ln{x_1} \ln{x_2}
+ \left (- \frac{29}{8}+ \ln{r_1} + \frac{15}{8} \ln{\eta} \right ) \ln{x_1}
+ \left (- \frac{25}{16} + \ln{r_2} + \frac{9}{8} \ln{\eta} \right) \ln{x_2} \non
&&\hspace{-2cm}+ \frac{1}{2} \ln{r_1} - \frac{1}{4} \ln^{2}{r_1} + \ln{r_2}
- \frac{9}{8} \ln{\eta} - \frac{1}{8} \ln^{2}{\eta} + \frac{37 \pi^2}{32}
+ \frac{91}{32} \Biggr ],\label{eq:ffnlot3}
\eeq
where $r_i=m_B^2/\xi_i^2$ with the choice of $\xi_1=25 m_B$\cite{prd85-074004}.
According to the analytical and numerical evaluations in Ref.~\cite{cheng14a},
we get to know that the NLO twist-2 and NLO twist-3 contribution to the form factor
of $B \to \pi$ transition are similar in size but have an opposite sign,
which leads to a strong cancelation between them and consequently
results in  a small total NLO contribution, $\sim 7\%$ variation to the full
LO pQCD prediction for the case of $f^+(q^2)$ in the range of
$0\leq q^2 \leq 12$ GeV$^2$, as illustrated explicitly in Fig.~8 of
Ref.~\cite{cheng14a}.

In this paper we adopt the above NLO factors $F^{(1)}_{\rm T2}(x_i,\mu,\mu_f,q^2)$ and
$F^{(1)}_{\rm T3}(x_i,\mu,\mu_f,q^2)$ directly, and then
extend the expressions to the case for $B \to K$ transition under the
assumption of $SU(3)$ flavor symmetry, by making the proper replacements,
such as $r_\pi=m_\pi/m_B \to r_k=m_k / \mbs$, $m_B \to \mbs$ and
$\phi_\pi^{A,P,T} \to \phi_K^{A,P,T}$, for the expressions as
given in Eqs.~(\ref{eq:ffnlot2},\ref{eq:ffnlot3}).

%%---------------------------------------------------------------

\subsection{NLO decay amplitudes}\label{sec:nlo2}

For the sake of comparison and convenience we denote all currently known NLO
contributions except for those NLO twist-2 and twist-3 contributions to the
form factors by the label ``Set-A", as described in previous subsection.
For the four considered $B_s^0 \to K\pi, KK$ decays, the Set-A NLO contributions
will be included in a simple way:
\beq
{\cal A}_{K \pi} &\to & {\cal A}_{K\pi}
+ \sum_{q=u,c,t} V_{qb}V_{qd}^*\;{\cal M}_{K^+ \pi^-}^{(QL)}
+ V_{tb}V_{td}^*\; {\cal M}_{K^+\pi^-}^{(MP)}, \label{eq:nlo01} \\
{\cal A}_{K K} &\to & {\cal A}_{K K}
+ \sum_{q=u,c,t} V_{qb}V_{qs}^*\; {\cal M}_{K^+ K^-}^{(QL)}
+ V_{tb}V_{ts}^* \; {\cal M}_{K^+ K^-}^{(MP)}, \label{eq:nlo03}
\eeq
where the quark-loop and magnetic penguin amplitudes
${\cal M}^{(QL)}_{XY} $ and ${\cal M}^{(MP)}_{XY}$ are of the form
\beq
{\cal M}^{(QL)}_{K^+ \pi^-}&=& -8m_{B_s}^4\frac{{C^2_F}}{\sqrt{6}}
\int_0^1 dx_1dx_2dx_3 \int_0^\infty b_1db_1b_3db_3 \;\phi_{B_s}(x_1) \non
&& \hspace{-1cm}\times
\Biggl \{ \Bigl [ (1+x_3)\phi_{\pi}^A(x_2) \phi_K^A(x_3)
+ r_K(1-2x_3)\phi^A_{\pi}(x_2)(\phi_K^P(x_3)+\phi_K^T(x_3))  \non
&&\hspace{-1cm}
+ 2r_{\pi}\phi_{\pi}^P(x_2) \phi_K^A(x_3) \Bigr ]
\cdot \alpha_s^2(t_a)\cdot h_e(x_1,x_3,b_1,b_3)\cdot
\exp[-S_{ab}(t_a)]\cdot C^{(q)}(t_a,l^2)\non
&&
+ \hspace{-1cm}2r_K\phi_{\pi}^A(x_2)\phi_K^P(x_3)
\cdot \alpha_s^2(t_b)\cdot h_e(x_3,x_1,b_3,b_1)
 \exp[-S_{ab}(t_b)]\cdot C^{(q)}(t_b,l'^2)\Biggr \},
\label{eq:mqkpi}
\eeq
\beq
{\cal M}^{(MP)}_{K^+\pi^-}  &=&-16m_{B_s}^6\frac{{C^2_F}}{\sqrt{6}}
\int_0^1 dx_1dx_2dx_3 \int_0^\infty b_1db_1b_2db_2b_3db_3\; \phi_{B_s}(x_1) \non
&& \cdot \Biggl \{ \Bigl [ (1-x_3) \phi_{\pi}^A(x_2)
\left [ 2\phi_K^A(x_3) + r_K (3+x_3)\phi_K^P(x_3)
+ r_K(1-x_3) \phi_K^T(x_3) \right ]  \non
&& - r_{\pi}x_2(1+x_3)\left ( 3\phi_{\pi}^P(x_2) -\phi_{\pi}^T(x_2) \right )
\phi_K^A(x_3) \Bigr]\non
&& \cdot \alpha_s^2(t_a)\cdot h_g(x_i,,b_i)\cdot
\exp[-S_{cd}(t_a)] \cdot C_{8g}^{eff}(t_a)  \non
&&
+ 4r_K\phi_{\pi}^A(x_2)\phi_K^P(x_3)\cdot
\alpha_s^2(t_b) h'_g(x_i,b_i) \cdot \exp[-S_{cd}(t_b)]
\cdot C_{8g}^{eff}(t_b)\Biggr \},
\label{eq:mgkpi} \\
\sqrt{2}{\cal M}^{(QL)}_{K^0 \pi^0}&=& {\cal M}^{(QL)}_{\pi^- K^+},\qquad
\sqrt{2}{\cal M}^{(MP)}_{K^0 \pi^0}= {\cal M}^{(MP)}_{\pi^- K^+},\label{eq:nlo00}
\eeq
\beq
{\cal M}^{(QL)}_{K^+ K^-}&=& -8m_{B_s}^4\frac{C^2_F}{\sqrt{6}}
\int_0^1 dx_1dx_2dx_3 \int_0^\infty b_1db_1b_3db_3 \; \phi_{B_s}(x_1) \non
&& \hspace{-1cm}\times
\Biggl \{ \Bigl [(1+x_3)\phi_K^A(x_2) \phi_K^A(x_3)
+ r_K (1-2x_3)\phi_K^A(x_2) \left ( \phi_K^P(x_3)+\phi_K^T(x_3) \right )\non
&&\hspace{-1cm} + 2r_{k}\phi_K^P(x_2)\phi_K^A(x_3) \Bigr]
\cdot \alpha_s^2(t_a)\cdot  h_e(x_1,x_3,b_1,b_3)
\cdot \exp[-S_{ab}(t_a)] \cdot C^{(q)}(t_a,l^2) \non
&& \hspace{-1cm}
+ 2r_K\phi_k^A(x_2)\phi_K^P(x_3)\cdot \alpha_s^2(t_b) \cdot h_e(x_3,x_1,b_3,b_1)
\cdot \exp[-S_{ab}(t_b)] \cdot C^{(q)}(t_b,l'^2)\Biggr\},
\label{eq:mqkk}
\eeq
\beq
{\cal M}^{(MP)}_{K^+ K^-} &=& - 16m_{B_s}^6\frac{C^2_F}{\sqrt{6}}
\int_0^1 dx_1dx_2dx_3 \int_0^\infty b_1db_1b_2db_2b_3db_3\; \phi_{B_s}(x_1) \non
&& \hspace{-1cm}\cdot
\Biggl \{ \Bigl[ (1-x_3) \left [2\phi_K^A(x_3) + r_K(3\phi_K^P(x_3)
+\phi_K^T(x_3)) +r_Kx_3(\phi_K^P(x_3)-\phi_K^T(x_3)) \right] \phi_K^A(x_2)\non
&& \hspace{-1cm}- r_K x_2(1+x_3)\left ( 3\phi_K^P(x_2)-\phi_K^T(x_2) \right ) \phi_K^A(x_3)
\Bigr ]\non
&& \hspace{-1cm} \cdot
\alpha_s^2(t_a)\cdot h_g(x_i,b_i) \cdot \exp[-S_{cd}(t_a)]\cdot
C_{8g}^{eff}(t_a) \non
&& \hspace{-1cm}
+ 4r_K \phi_K^A(x_2)\phi_K^P(x_3) \cdot
\alpha_s^2(t_b)\cdot h'_g(x_i,b_i) \cdot \exp[-S_{cd}(t_b)]
\cdot C_{8g}^{eff}(t_b)\Biggr \}, \label{eq:mgkk} \\
{\cal M}^{(QL)}_{\bar K^0 K^0} &=& {\cal M}^{(QL)}_{K^+ K^-}, \qquad
{\cal M}^{(MP)}_{\bar K^0 K^0} = {\cal M}^{(MP)}_{K^+ K^-},
\label{eq:mpkk}
\eeq
where the terms proportional to $r_\pi r_K$ or $r_K^2$ are not shown for the sake of
simplicity. The functions $h_e, h_g$ and $h_g^\prime$, the hard scales $t_a$ and
$t_b$, as well as the Sudakov factors $S_{ab}(t)$ and $S_{cd}(t)$ in
Eqs.(\ref{eq:mqkpi}-\ref{eq:mgkk}) will be given in Appendix A.

\section{Numerical results}\label{sec:n-d}

In the numerical calculations the following input parameters will be used.
\beq
\Lambda_{\overline{\mathrm{MS}}}^{(5)} &=& 0.225 {\rm GeV},
\quad  f_{B_{s}} = (0.23\pm 0.02) {\rm GeV}, \quad  f_K = 0.16  {\rm GeV},
\quad   f_{\pi} = 0.13{\rm GeV},\non
M_{B_{s}} &=&  5.37 {\rm GeV},\quad   m_K=0.494{\rm GeV},\quad
m_0^\pi = 1.4 {\rm GeV}, \quad   m_0^K = 1.9 {\rm GeV}, \non
\tau_{B_s^0} &=& 1.497 {\rm ps}, \quad m_b=4.8 {\rm GeV},
\quad M_W = 80.42 {\rm GeV}. \label{eq:para}
\eeq
For the CKM matrix elements, we also take the same values
as being used in Ref.~\cite{ali07}, and neglect the small errors on
$V_{ud}, V_{us}$, $V_{ts}$ and $V_{tb}$
\beq
|V_{ud}|&=& 0.974, \quad |V_{us}|=0.226,
\quad |V_{ub}|=\left ( 3.68^{+0.11}_{-0.08}\right)\times 10^{-3},\non
\quad |V_{td}|&=&\left ( 8.20^{+0.59}_{-0.27}\right)\times 10^{-3},\quad
|V_{ts}|= 40.96\times 10^{-3}, \quad |V_{tb}|= 1.0,\non
\alpha&=&(99^{+4}_{-9.4})^\circ ,\quad
\gamma=(59.0^{+9.7}_{-3.7})^\circ, \quad
\arg\left [-V_{ts}V_{tb}^* \right] =1^\circ.
\label{eq:angles}
\eeq

\subsection{Branching Ratios}

For the considered $B_s^0$ decays, the decay amplitude for a given decay mode
with $b \to d,s$ transitions can be generally written as
\beq
{\cal A}(\bar{B}_s^0 \to f)|_{b\to d} &=& V_{ub}V_{ud}^* T -V_{tb}V_{td}^*  P
= V_{ub}V_{ud}^*T  \left [ 1 + z e^{ i ( -\alpha + \delta ) } \right],\label{eq:ma}\\
{\cal A}(\bar{B}_s^0 \to f)|_{b \to s} &=& V_{ub}V_{us}^* T^\prime -V_{tb}V_{ts}^*  P^\prime
=V_{ub}V_{us}^* T^\prime  \left [ 1 + z^\prime e^{ i ( \gamma + \delta^\prime ) } \right],
\label{eq:ma2}
\eeq
where $\alpha $ and $\gamma$ are the weak phase ( the CKM angles ),
$\delta=\arg[P/T]$ and $\delta^\prime=\arg[P^\prime/T^\prime]$  are the relative strong phase between the tree (T)
and penguin (P) diagrams, and the parameter  ``z"  is the ratio of penguin to
tree contributions with the definition
\beq
z=\left|\frac{V_{tb} V_{td}^*}{ V_{ub}V_{ud} ^*} \right| \left|\frac{P}{T}\right|,
 \quad
z^\prime=\left|\frac{V_{tb} V_{ts}^*}{ V_{ub}V_{us} ^*}
\right| \left|\frac{P^\prime}{T^\prime}\right|.
\label{eq:zz}
\eeq
The ratio $z$ and the strong phase $\delta$ can be calculated in the
pQCD approach. The CP-averaged branching ratio, consequently, can be
defined as
\beq
{\rm Br}(\bar{B}_s^0\to f) = \frac{G_F^2 \tau_{B_s^0}}{32\pi m_B} \;
\frac{1}{2} \left [ |{\cal A}(\bar{B}_s^0\to f)|^2
+|{\cal A}(B_s^0\to \bar{f})|^2\right],
\label{eq:br0}
\eeq
where $\tau_{B_s^0}$ is the lifetime of the $B_s^0$ meson.

\begin{table}
\caption{ The pQCD predictions for the branching ratios  ( in units of $10^{-6}$)
of the four $B_s ^0\to K\pi, KK$ decays. The label ``LO"  and ``NLO" means the
leading order and the full next-to-leading order pQCD predictions,
while ``Set-A" means only NLO twist-2 and twist-3 contributions to form factors
are not taken into account.
The values listed in the fifth, sixth and seventh column are the LO pQCD predictions
\cite{ali07}, the QCDF predictions \cite{npb675}, and currently available data
\cite{hfag2012,pdg2012}. }
\label{tab:br1}
\begin{tabular*}{16cm}{@{\extracolsep{\fill}}l|c|ll|ll|ll} \hline\hline
 Mode & Class & LO & pQCD \cite{ali07} & Set-A &  NLO    & QCDF \cite{npb675}& Data \\ \hline
 $\bar B_s^0\to K^+\pi^-   $ &T &$7.30 $&$7.6^{+3.3}_{-2.5}   $ &$6.4$&$5.7^{+2.2+0.5+0.2}_{-1.7-0.6-0.3}     $&$10.2^{+6.0}_{-5.2}  $&$5.4\pm 0.6   $\\
 $\bar B_s^0\to K^0\pi^0   $ &C &$0.19 $&$0.16^{+0.12}_{-0.07}$ &$0.30$&$0.28^{+0.10+0.03+0.02}_{-0.06-0.02-0.01}$&$0.49^{+0.62}_{-0.35}$&        \\
 $\bar B_s^0\to K^+K^-     $ &P &$13.1 $&$13.6^{+8.6}_{-5.2}  $ &$20.3$&$19.7^{+6.2+2.4+0.2}_{-4.8-2.2-0.2}      $&$22.7^{+27.8}_{-13.0}$&$24.5\pm 1.8  $\\
 $\bar B_s^0\to\bar{K}^0K^0$ &P &$13.3 $&$15.6^{+9.7}_{-6.0}  $ &$21.2$&$20.2^{+6.5+2.4+0.0}_{-4.9-2.2-0.0}      $&$24.7^{+29.4}_{-14.0}$&        \\
\hline\hline
 \end{tabular*}
\end{table}

In Table I, we list the pQCD predictions for the branching ratios
of the four $B_s^0 \to K\pi, KK$ decays. The label ``LO" and ``NLO" means the
pQCD predictions at the leading order and with the inclusion of all currently known
NLO contributions. The label``Set-A" means the pQCD predictions without
the inclusion of the newly known NLO twist-2 and twist-3 contributions to
the form factors of $B \to K$ transitions.
For the sake of comparison, we also show the LO pQCD predictions as given in
Ref.~\cite{ali07} in the fourth column, and list the NLO theoretical predictions
obtained by employing the QCD factorization approach as given in Ref.~\cite{npb675}
in the seventh column.
The corresponding errors of the previous LO pQCD predictions \cite{ali07} 
and the QCDF predictions\cite{npb675} are the combined total errors.
The currently available experimental measurements \cite{hfag2012,pdg2012}
are also shown in the last column of Table I.

The theoretical errors of the NLO pQCD predictions as shown in the sixth column of
Table I are induced by the uncertainties of the input parameters.
The first dominant error comes from $\omega_b=0.50 \pm 0.05$ and
$f_{B_s}=0.23\pm 0.02$ GeV, added in quadrature.
The second error arises from the uncertainties of the CKM matrix
elements $|V_{ub}|$ and $|V_{cb}|$, as well as the CKM angles
$\alpha$ and $\gamma$ as given in Eq.~(\ref{eq:angles}).
The third error comes from the uncertainties of
relevant Gegenbauer moments: $a_1^K=0.17\pm 0.05$, $a_2^K=0.20\pm
0.06$ and $a_2^\pi=0.44^{+0.10}_{-0.20}$, added in quadrature again.
We here assigned roughly a $30\%$ uncertainty for Gegenbauer moments
to estimate the resultant errors for the pQCD predictions of the branching ratios.

From the numerical results of the branching ratios, we have the following observations:
\begin{enumerate}
\item
The LO pQCD predictions for the branching ratios as given in Ref.~\cite{ali07} are confirmed by
our independent calculations.
The small differences between the values in column three and four
are mainly induced by the different choices of the scales $\Lambda_{QCD}^{(4)}$
and $\Lambda_{QCD}^{(5)}$:
we take $\Lambda_{QCD}^{(5)}=0.225$ GeV and $\Lambda_{QCD}^{(4)}=0.287$ GeV,
instead of the values of $\Lambda_{QCD}^{(5)}=0.193$ GeV and $\Lambda_{QCD}^{(4)}=0.25$ GeV
as being used in Ref.~\cite{ali07}.

\item
The NLO contributions can interfere with the LO part constructively or
destructively for different decay modes. The inclusion of NLO contributions 
can leads to a better agreement between the theoretical 
predictions and currently available measured values.

\item
The $\bar{B}^0_s \to K^+ \pi^-$ decay is a ``tree" dominated decay mode,
the NLO contribution leads to a $22\%$ decrease to the LO pQCD prediction only.
For other three ``Color-suppressed" and "QCD-penguin" decay modes,
however, the NLO contribution leads to a $\sim 50\%$ enhancement to the LO ones,
which in turn play an important role in interpreting the observed large branching ratio 
$Br(B_s^0\to K^+K^-)=(24.5 \pm 1.8)\times 10^{-6}$ \cite{hfag2012,pdg2012}.

\end{enumerate}

\subsection{CP-violating asymmetries}

Now we turn to the evaluations of the CP-violating asymmetries of
the considered four $B^0_s$ decays in the pQCD approach.
For $B_s^0\to K^\mp \pi^\pm$ decays, the definition for its direct
CP violating asymmetry is very simple\cite{lhcb1}.
For neutral $B_s^0$ decays into a CP eigenstate $\bar{f}=\eta_{CP} f$
with $\eta_{CP}=\pm 1$ for the CP-even and CP-odd final states,
the time dependent CP asymmetry can be defined as \cite{lhcb2,prd52}
\beq
\mathcal{A}(t)=\frac{\Gamma_{\bar{B}^0_s \to f }(t)-\Gamma_{B^0_s \to f }(t)}{
\Gamma_{\bar{B}^0_s \to f}(t)+\Gamma_{B^0_s \to f}(t)}
=\frac{\cala_f \cos(\Delta m_s\; t) + \cals_f \sin(\Delta m_s\; t)}{
\cosh\left(\frac{\Delta\Gamma_s}{2} t\right)
+ \calh_f \sinh\left(\frac{\Delta\Gamma_s}{2} t\right)},
\label{eq:cpv-def1}
\eeq
where $\Delta m_s$ and $\Delta\Gamma_s$ are the mass and width differences
of the $B^0_s-\bar{B}^0_s$ system mass eigenstates.
The direct CP violating asymmetry $\cala_f$, the mixing-induced CP violating
asymmetry  $\cals_f$ and $\calh_f$ are defined as in Refs.~\cite{lhcb2,prd52}:
\beq
\cala_f = \frac{|\lambda_f|^2-1}{1+|\lambda_f|^2},\quad
\cals_f = \frac{2 {\rm Im} \lambda_f}{1+|\lambda_f|^2},\quad
\calh_f = \frac{2 {\rm Re} \lambda_f}{1+|\lambda_f|^2}, \label{eq:cpv-def2}
\eeq
where the three factors satisfy the normalization relation:
$|\cala_f|^2 + |\cals_f|^2 +|\calh_f|^2 =1$, and the CP-violating parameter
$\lambda_f$ is defined as
\beq
\lambda_f = \frac{q}{p}\frac{\bar{A}_f}{A_f}
= \eta_f \;e^{2i\epsilon}\frac{A(\bar B_s \to f)}{A(B_s \to f)}
\label{eq:lambdaf}
\eeq
where $\epsilon =\arg[-V_{ts}V_{tb}^*]$ is very small in size and can be
neglected safely. It is worth of mentioning that the parameter $\cala_f $
and $\calh_f$ defined in Eqs.~(\ref{eq:cpv-def1},\ref{eq:cpv-def2})
have opposite sign with the parameter ${\cal C}_f$  and $\cala_f^{\Delta \Gamma}$
as defined in Ref.\cite{lhcb2} : i.e., $\cala_f = - {\cal C}_f$ and
$\calh_f = - \cala_f^{\Delta \Gamma}$.

In Table II and III, we list the pQCD predictions (in unit of $10^{-2}$) for
the direct CP-violating asymmetry $\cala_f$, the mixing-induced CP-violating asymmetry
$\cals_f$ and $\calh_f$ of the considered $B_s^0$ decays, respectively.
As a comparison, the LO pQCD predictions as given in Ref.~\cite{ali07},
the QCDF predictions as given in Ref.~\cite{npb675,sun2003} and the measured
values \cite{lhcb1,lhcb2} are listed in Table \ref{tab:acp1} and Table \ref{tab:acp2}.
The errors of our NLO pQCD predictions for CP-violating asymmetries
are defined in the same way as those for the branching ratios.

\begin{table}
\caption{The same format as in Table I, but for the pQCD predictions (in unit of $10^{-2}$ )
for the direct CP asymmetries $\cala_f$.
The previous LO pQCD predictions \cite{ali07}, the QCDF predictions \cite{npb675}
and the measured values \cite{lhcb1,lhcb2} are also listed. }
\label{tab:acp1}
\begin{tabular*}{16cm}{@{\extracolsep{\fill}}l|c|ll|ll|ll} \hline\hline
 Mode &                     Class & LO     & pQCD \cite{ali07}  & Set-A & NLO                                        & QCDF \cite{npb675} & Data  \\ \hline
 $\bar B_s^0\to K^+\pi^-  $    &$T $&$27.6  $&$24.1^{+5.6}_{-4.8}  $&$36.2 $&$38.7^{+5.0+2.1+2.2}_{-5.0-1.8-1.8}        $&$-6.7^{+15.6}_{-15.3}$& $27\pm 4$\cite{lhcb1} \\
 $\bar B_s^0\to K_S^0\pi^0  $  &$C $&$62.9  $&$59.4^{+7.9}_{-12.5} $&$84.8 $&$83.0^{+5.8+3.4+2.3}_{-5.6-2.6-2.7}        $&$42^{+47}_{-56}      $&       \\
 $\bar B_s^0\to K^+K^-    $    &$P $&$-13.7 $&$-23.3^{+5.0}_{-4.6} $&$-17.1$&$-16.4^{+0.3+0.6+0.6}_{-0.1-0.4-0.6} $&$4.0^{+10.6}_{-11.6} $&$-14\pm 12$\cite{lhcb2} \\
 $\bar B_s^0\to \bar K^0 K^0$  &$P $&$0     $&$0                   $&$-0.7$&$-0.7\pm 0.1 $&$0.3\pm 0.1          $&   \\
\hline\hline
\end{tabular*}
\end{table}

\begin{table}
\caption{The same format as in Table I, but for the pQCD predictions (in unit of $10^{-2}$ ) for the
mixing-induced CP asymmetries  $\cals_f$ and $\calh_f$ (the second row).
The previous LO pQCD predictions \cite{ali07}, the QCDF predictions \cite{sun2003} and the measured
values \cite{lhcb1,lhcb2} are also listed. }
\label{tab:acp2}
\begin{tabular*}{16cm}{@{\extracolsep{\fill}}l|c|ll|ll|ll} \hline\hline
 Mode &                     Class & LO     & pQCD \cite{ali07}  & Set-A & NLO                   &    QCDF\cite{sun2003} & Data\cite{lhcb2}  \\ \hline
$\bar B_s^0\to K_S^0\pi^0$  &C &$-56.2 $&$-61^{+24}_{-20}$&$-50.0$&$-52.9^{+8.0+4.2+4.7}_{-8.2-4.3-4.4}            $&$45  $&$$  \\
                            &  &$-53.7 $&$-52^{+23}_{-17}$&$-17.8$&$-17.4^{+0.9+2.0+4.8}_{-0.1-1.0-4.1}          $&$-   $&$$  \\
$\bar B_s^0\to K^+K^-$      &P &$37.1  $&$28^{+5}_{-5}   $&$22.0 $&$20.6^{+1.9+1.4+0.8}_{-1.8-1.3-0.7}     $&$27  $&$30\pm 13$ \\
                            &  &$92.0  $&$93 ^{+3}_{-3}  $&$96.0 $&$96.5^{+0.3+0.1+0.1}_{-0.4-0.2-0.2} $&$-   $&$$ \\
$\bar B_s^0\to \bar{K}^0K^0$&P &$ -    $&$4              $&$-0.2 $&$-0.2                               $&$-3.5$&$$ \\
                            &  &$100   $&$\sim 100            $&$\sim 100$&$\sim 100$ & $-$ & \\
                            \hline \hline
\end{tabular*}
\end{table}

From the pQCD predictions and currently available data for the CP violating
asymmetries of the considered $\bar{B}_s^0$ decays, we find that
(a) the LO pQCD predictions obtained in this paper agree well with those as given in
Ref.~\cite{ali07}; (b) For the CP-violating asymmetries of the considered $\bar{B}_s^0$ decays,
the effects of the NLO contributions are small or moderate in size; and (c)
for $\bar{B}_s^0 \to K^\pm \pi^\mp$ and $\bar{B}_s^0\to K^+ K^-$ decays,
the pQCD predictions for both $\cala_f$ and $\cals_f$ agree well with those measured values
in both the sign and the magnitude.

\section{Summary}

In this paper, we calculated the branching ratios and CP-violating asymmetries of the four
$\bar B_s^0\to K\pi, KK$ decays, with the inclusion of all known NLO contributions, especially
the NLO twist-2 and twist-3 contributions to the form factors to $B_s \to K$ transition.
From our calculations and phenomenological analysis, we found the following results:
\begin{enumerate}
\item
For the considered four decays, the NLO contribution can interfere with the
LO part constructively or destructively for different decay modes. 
The currently available data can be interpreted by the inclusion of 
the NLO contribution.  

\item
For $Br(\bar{B}^0_s \to K^+ \pi^-)$, the NLO contribution leads to a $22\%$ decrease
to the LO pQCD prediction.
For other three decay modes, however, the NLO contributions can provide $\sim 50\%$ 
enhancements to the LO ones and therefore play an important role in interpreting 
the observed large branching ratio 
$Br(\bar{B}_s^0\to K^+K^-)=(24.5 \pm 1.8)\times 10^{-6}$.

\item
For the CP-violating asymmetries, the effects of the NLO contributions are small or moderate
in size. For $\bar{B}_s^0 \to K^+ \pi^-$ and $\bar{B}_s^0 \to K^+K^- $ decays, 
the pQCD predictions for the direct and mixing induced CP-violating asymmetries 
agree very well with the measured values in both the sign and the magnitude.

\end{enumerate}

%%=================================================

\begin{acknowledgments}
This work is partly supported by the National Natural Science
Foundation of China under Grant No.11235005.

\end{acknowledgments}

%%%%%%%%%%%%%%%%%%%%%%%%%%%%%%%%%%%%%%%%%%%%%%%%%%%%%%%%%%%%%%%%%%%%%%%%%%%%%%%%%%
%                                        Appendix
%%%%%%%%%%%%%%%%%%%%%%%%%%%%%%%%%%%%%%%%%%%%%%%%%%%%%%%%%%%%%%%%%%%%%%%%%%%%%%%%5

\begin{appendix}

\section{Related hard functions and Sudakov factors}\label{sec:aa}

We here list the hard function $h_i$ and the Sudakov factors $S_{ab}(t)$ and $S_{cd}(t)$
appeared in the expressions of the decay amplitudes in Eqs.~(\ref{eq:mqkpi}-\ref{eq:mgkk}).
The hard functions $h_i(x_j,b_j)$ are obtained by making the Fourier
transformations of the hard kernel $H^{(0)}$.
\beq
h_e(x_1,x_3,b_1,b_3)&=& \Bigl [ \theta(b_1-b_3)I_0\left(\sqrt{x_3}
\mbs b_3 \right )K_0\left (\sqrt{x_3} \mbs b_1\right )
 +\theta(b_3-b_1)I_0\left (\sqrt{x_3}  \mbs  b_1\right ) \non
&& \cdot K_0\left (\sqrt{x_3}\mbs  b_3 \right )\Bigr ]
\cdot K_0\left (\sqrt{x_1 x_3} \mbs  b_1\right ) S_t(x_3), \label{he1}
\eeq
\beq
h_g(x_i,b_i)&=&-\frac{i\pi}{2}S_t(x_3) \Bigl[ J_0\left (\sqrt{x_2\bar{x}_3} \mbs b_2\right )
+iN_0\left (\sqrt{x_2\bar{x}_3} \mbs b_2\right )\Bigr ] \cdot K_0\left (\sqrt{x_1x_3}M_{B_s}b_1\right )
\non
&& \hspace{-1.5cm}
\cdot \int_0^{\pi/2}\!d\theta \tan{\theta}
\cdot J_0\left(\sqrt{x_3} \mbs b_1\tan{\theta}\right )
J_0\left (\sqrt{x_3} \mbs b_2\tan{\theta}\right )
\cdot J_0\left (\sqrt{x_3} \mbs b_3\tan{\theta}\right ),
\eeq
\beq
h'_g(x_i,b_i)&=& -S_t(x_1) K_0\left (\sqrt{x_1x_3} \mbs b_3 \right )
\cdot \int_0^{\pi/2}\!d\theta \tan{\theta}\cdot
J_0\left (\sqrt{x_1} \mbs b_1\tan{\theta}\right )
\non &&
\ \ \ \ \cdot J_0\left (\sqrt{x_1} \mbs b_2\tan{\theta} \right )
\; J_0\left (\sqrt{x_1} \mbs b_3\tan{\theta} \right )\non
&& \ \ \ \ \times
\left\{ \begin{array}{ll}\frac{i\pi}{2}
\left [ J_0\left (\sqrt{x_2-x_1} \mbs b_2 \right )
+iN_0\left (\sqrt{x_2-x_1} \mbs b_2\right )\right ],& x_1<x_2,\\
K_0\left (\sqrt{x_1-x_2} \mbs b_2 \right ),& x_1>x_2,\end{array}\right.
\eeq
with $K_0$, $I_0$ and $J_0$ are the Bessel functions~\cite{isg}. And the threshold
resummation form factor $S_t(x_i)$ can be found in Ref.~\cite{kls01}.

The Sudakov factors appeared in Eqs.~(\ref{eq:mqkpi}-\ref{eq:mgkk}) are defined as
\beq
S_{ab}(t) &=& s\left ( x_1\frac{ \mbs }{\sqrt 2},\,b_1 \right )
+ s\left (x_3\frac{ \mbs }{\sqrt{2}},\,b_3 \right )
+ s\left (\bar{x}_3\frac{ \mbs }{\sqrt 2},\,b_3 \right )
\non &&+\frac{5}{3}\int_{1/b_1}^t d\mu
\frac{\gamma_q(\alpha_s(\mu))}{\mu}+ 2\int_{1/ b_3}^t d\mu
\frac{\gamma_q(\alpha_s(\mu))}{\mu},
\label{eq:Sab}
\eeq
\beq
S_{cd}(t) &=& s\left ( x_1\frac{ \mbs }{\sqrt 2},\,b_1 \right )
+ s\left ( x_2\frac{ \mbs}{\sqrt{2}},\,b_2 \right )
+ s\left (\bar{x}_2 \frac{ \mbs}{\sqrt2},\,b_2\right )
+ s\left (x_3\frac{\mbs }{\sqrt{2}},\,b_1 \right )
\non &&
+ s\left (\bar{x}_3\frac{ \mbs}{\sqrt 2},\,b_1\right ) +\frac{11}{3}\int_{1/b_1}^t d\mu
\frac{\gamma_q(\alpha_s(\mu))}{\mu}+ 2\int_{1/ b_2}^t d\mu
\frac{\gamma_q(\alpha_s(\mu))}{\mu},
\label{eq:Scd}
\eeq
where $\bar{x}_i=1-x_i$, the function $s(Q,b)$ can be found in Refs.~\cite{plb555,epjc695}.
The hard scales $t_a$ and $t_b$ appeared in Eqs.~(\ref{eq:mqkpi}-\ref{eq:mgkk}) take the form of
\beq
t_a &=& {\rm max}\left\{\sqrt{x_1 x_3}\mbs,
         \sqrt{x_3} \mbs, \sqrt{x_2(1-x_3)}\mbs, 1/b_1,1/b_3 \right\},\non
t_b &=& {\rm max}\left \{\sqrt{x_1 x_3} \mbs, \sqrt{x_1} \mbs,
         \sqrt{| x_1-x_2|} \mbs,1/b_1,1/b_3 \right\},
\eeq
where the energy scale $\sqrt{x_2(1-x_3)}\mbs$ and $\sqrt{| x_1-x_2|} \mbs$ come from the invariant mass
of the gluon $l^2=x_2(1-x_3)\mbs^2$ and $l^{\prime 2}=(x_1-x_2) \mbs^2$. They are chosen as the maximum
energy scale appearing in each diagram to kill the large logarithmic radiative corrections.

\end{appendix}

%%%%%%%%%%%%%%%%%%%%%%%%%%%%%%%%%%%%%%%%%%%%%%%%%%%%%%%%%%%%%%%%%%%%%%%%%%%%%%%%%%%%%%%%%%%%%%5
%                                 reference
%%%%%%%%%%%%%%%%%%%%%%%%%%%%%%%%%%%%%%%%%%%%%%%%%%%%%%%%%%%%%%%%%%%%%%%%%%%%%%%%%%%%%%%%%%%%%%%%%

%\newpage


\begin{thebibliography}{99}

\bibitem{lhcb1}
R.~Aaij {\it et al.} (LHCb Collaboration),  \prl {\bf 108}, 201601 (2012);
%% First Evidence of Direct CPV in Charmless Two-Body Decays of B^0_s Mesons
R. Aaij et al. (LHCb collaboration), \prl {\bf 110}, 221601 (2013).
%% First Observation of CP Violation in the Decays of B^0_s Mesons

\bibitem{lhcb2}
R. Aaij et al. (LHCb collaboration), \jhep {\bf 10}, 183 (2013);
%% First measurement of time-dependent CP violation in $B_s \to K^+K^-$ decays
\plb {\bf 716}, 393(2012). 	
%% Measurement of the effective $B_s\to K^+K^-$ lifetime

\bibitem{bspipi}
Y.~Li, C.D.~L\"u, Z.J.~Xiao, and X.Q.~Yu,  \prd {\bf 70}, 034009 (2004);
%% Branching ratio and CP asymmetry of $B_s^0 \to \pi^+\pi^-$ decays in the pQCD approach
X.Q.~Yu, Y.~Li, and C.D.~L\"u, \prd {\bf 71}, 074026 (2005);
%% Branching ratio and CP violation of B_s\to \pi K decays in the pQCD approach
\prd {\bf 73}, 017501 (2006);
%% Study of B_s \to \pi \rho decays in the pQCD approach
J.~Zhu, Y.L.~Shen, and C.D.~L\"u, \jpg {\bf 32}, 101  (2006).
%% B_s\to \rho (\omega)K^* with the perturbative QCD approach

\bibitem{pieta}
Z.J. ~Xiao, X.~Liu and H.S.~Wang, \prd {\bf 75}, 034017 (2007).
%% Br and CP asymmetry of B_s \to \pi \eta decays in the perturbative QCD approach

\bibitem{ali07}
A.~Ali, G.~Kramer, Y.~Li, C.D.~L\"u, Y.L.~Shen, W.~Wang and Y.M.~Wang,
\prd {\bf 76}, 074018 (2007).
%% Charmless nonleptonic B_s decays to PP, PV, and VV final states in the pQCD approach

\bibitem{xiao08a}
J.Liu, R.Zhou and Z.J. Xiao, arXiv:0812.2312[hep-ph].
%% B_s \to PP decays and the NLO contributions in the pQCD

\bibitem{chenbs99}
Y.H.~Chen, H.Y.~Cheng, B.~Tseng,  \prd {\bf 59}, 074003 (1999).
%% Charmless hadronic two-body decays of B_s mesons

\bibitem{npb675}
M.~Beneke and M.~Neubert, \npb {\bf 675}, 333 (2003).
%% QCD factorization for B \to PP and B \to PV decays

\bibitem{sun2003}
J.F.~Sun, G.H.~Zhu, D.S.~Du, \prd {\bf 68}, 054003  (2003).
%% Phenomenological analysis of charmless decays B(s) \to PP, PV, with QCD factorization

\bibitem{chengbs09}
H.Y.~Cheng and C.K.~Chua, \prd {\bf 80}, 114026 (2009).
%% QCD factorization for charmless hadronic Bs decays revisited

\bibitem{xiao08b}
Z.J.~Xiao, Z.Q.~Zhang,  X.~Liu, and L.B.~Guo, \prd {\bf 78}, 114001 (2008).
%% Brs and CP asymmetries of B \to K\etap decays in the pQCD approach

\bibitem{fan2013}
Y.Y. Fan, W.F. Wang, S.Cheng, and Z.J.~Xiao,\prd {\bf 87}, 094003 (2013).
%% Anatomy of B \to K \etap  decays in different mixing schemes and
%% effects of NLO contributions in the perturbative QCD approach

\bibitem{xiao2014}
W.~Bai, M.~Liu, Y.Y.~Fan, W.F.~Wang, S.~Cheng, and Z.J.~Xiao, \cpc {\bf 38}, 033101(2014).
%% Revisiting K\pi puzzle in the pQCD factorization approach

\bibitem{nlo05}
H.N.~Li, S.~Mishima, A.I.~Sanda, \prd {\bf 72}, 114005 (2005).
%% Resolution to the B \to \pi K puzzle


\bibitem{prd85-074004}
H.N.~Li, Y.L.~Shen, and Y.M.~Wang, \prd {\bf 85}, 074004 (2012).
%% Next-to-leading-order corrections to B \to \pi form factors in kT factorization

\bibitem{cheng14a}
S.~Cheng, Y.Y.~Fan, X.Yu, C.D.~L\"u and Z.J.~Xiao, arXiv:1402.5501v2[hep-ph].
%% The NLO twist-3 contributions to $B \to \pi$ form factors in $k_{T}$ factorization


\bibitem{li2003}
H.N.~Li,  Prog. Part. $\&$ Nucl. Phys. {\bf 51}, 85 (2003) and references therein.
%% QCD Aspects of Exclusive B Meson Decays

\bibitem{buras96}
G.~Buchalla, A.J.~Buras, M.E.~Lautenbacher, \rmp {\bf 68}, 1125 (1996).
%% Weak decays beyond leading logarithms

\bibitem{pball98}
V.M.~Braun and I.E.~Filyanov , \zpc {\bf 48}, 239 (1990);
P.~Ball, V.M.~Braun, Y.~Koike, and K.~Tanaka, \npb {\bf 529}, 323 (1998);
P.~Ball, \jhep {\bf 01}, 010 (1999).

\bibitem{pball06}
V.M.~Braun and A.~Lenz, \prd {\bf 70}, 074020 (2004);
P.~Ball and A.~Talbot, \jhep {\bf 06}, 063 (2005);
P.~Ball and R.~Zwicky, \plb {\bf 633}, 289 (2006);
A.~Khodjamirian, Th.~Mannel, and M.~Melcher, \prd {\bf 70}, 094002 (2004).

\bibitem{wu2014}
X.G.~Wu and T.~Huang, arXiv:1312.1355[hep-ph].
%% Heavy and light meson wavefunctions

\bibitem{o8g2003}
S.~Mishima and A.I.~Sanda, Prog. Theor. Phys. {\bf 110}, 549 (2003).

\bibitem{bbns99}
M.~Beneke, G.~Buchalla, M.~Neubert and C.T.~Sachrajda, \prl {\bf 83}, 1914 (1999);
%% QCD Factorization for B \to \pi \pi Decays: Strong Phases and CP Violation
%% in the Heavy Quark Limit
\npb {\bf 591}, 313 (2000).
%% QCDf for exclusive non-leptonic B-meson decays: general arguments and the case
%%  of heavy¨Clight final states

\bibitem{kls01}
T.~Kurimoto, H.N.~Li, and A.I.~Sanda, \prd {\bf 65}, 014007 (2001);
%Leading power contributions to B ---> pi, rho transition form-factors%
C.D.~Lu and M.Z.~Yang, \epjc {\bf 28}, 515 (2003).
%% B to light meson transition form factors calculated in pQCD approach

\bibitem{hfag2012}
Y.~Amhis {\it et al}., (Heavy Flavor Averaging Group), arXiv:1207.1158 [hep-ex].

\bibitem{pdg2012}
J.~Beringer {\it et al.} (Particle Data Group),  \prd {\bf 86}, 010001 (2012).

\bibitem{prd52}
I.~Dunietz, \prd {\bf 52}, 3048 (1995).
%% B_s-\bar{B}_s mixing, CPV, and extraction of CKM phase from untagged Bs data

\bibitem{isg}
I.S.~Gradshteyn and I.M.~Ryzhik, Table of Integrals, Series, and Products,
Academic Press, 1980.

\bibitem{plb555}
H.N.~Li and K.Ukai, \plb{\bf 555}, 197(2003).
%%Threshold resummation for nonleptonic B meson decays

\bibitem{epjc695}
H.N.~Li and B.~Melic, \epjc {\bf 11}, 695 (1999).
%% Determination of heavy meson wave functions from B mesons

\end{thebibliography}
\end{document}